\pdfoutput=1
\documentclass[aps,prd,amsmath,floats,floatfix, twocolumn,
superscriptaddress,nofootinbib,showpacs,longbibliography]{revtex4-1}
\usepackage[T1]{fontenc}
\usepackage[utf8]{inputenc}
\usepackage{lmodern}
\usepackage{verbatim}
\usepackage[dvipsnames, usenames]{xcolor}
\definecolor{linkcolor}{rgb}{0.0,0.3,0.5}
\usepackage[hypertexnames=false, unicode, colorlinks=true, linkcolor=linkcolor,
citecolor=linkcolor, filecolor=linkcolor,urlcolor=linkcolor,
pdfusetitle]{hyperref}
\usepackage[all]{hypcap}
\usepackage{graphicx}
\usepackage{xspace}
\usepackage{amssymb}
\usepackage{microtype}

\usepackage[english]{babel}
\usepackage{blindtext}

\usepackage[normalem]{ulem} %for \sout
\usepackage{bm} % boldmath
\usepackage{mathrsfs}

\graphicspath{
	{figs/}
}

\DeclareMathAlphabet{\mathpzc}{OT1}{pzc}{m}{it}

\newcommand{\model}{\texttt{BHPTNRSur1dq1e4}} % shorter

\usepackage[nomessages]{fp}

\newcommand{\bhpt}{\texttt{BHPTNRSur1dq1e4}}
\newcommand{\seob}{\texttt{SEOBNRv4HM}}
\newcommand{\nrsur}{\texttt{NRHybSur3dq8}}
\newcommand{\nrrem}{\texttt{NRSur3dq8Remnant}}

\begin{document}
	
\title{Remnant black hole properties from numerical-relativity-informed perturbation theory and implications for waveform modelling}
\newcommand{\UMassDMath}{\affiliation{Department of Mathematics,
			University of Massachusetts, Dartmouth, MA 02747, USA}}
\newcommand{\UMassDPhy}{\affiliation{Department of Physics,
			University of Massachusetts, Dartmouth, MA 02747, USA}}
\newcommand{\CSCVRUMass}{\affiliation{Center for Scientific Computing and Visualization Research, University of Massachusetts, Dartmouth, MA 02747, USA}}
\newcommand{\URI}{\affiliation{Department of Physics, 
			University of Rhode Island, Kingston, RI 02881, USA}}    
	
\author{Tousif Islam}
\email{tislam@umassd.edu}
\UMassDPhy
\UMassDMath
\CSCVRUMass
	
\author{Scott E. Field}
\UMassDMath
\CSCVRUMass
	
\author{Gaurav Khanna}
\URI
\UMassDPhy
\CSCVRUMass
	
% Because hyperref only gets the *last* author, we need to be explicit.
\hypersetup{pdfauthor={Islam et al.}}
	
\date{\today}

%==========================================================================
\begin{abstract}
During binary black hole (BBH) mergers, energy and momenta are carried away from the binary system as gravitational radiation. Access to the radiated energy and momenta allows us to predict the properties of the remnant black hole. We develop a Python package \texttt{gw\_remnant} to compute the remnant mass, remnant spin, peak luminosity, and the final kick imparted on the remnant black hole from the gravitational radiation. Using this package, we compute the remnant properties of the final black hole in case of non-spinning BBH mergers with mass ratios ranging from $q=2.5$ to $q=1000$ using waveform modes generated from \texttt{BHPTNRSur1dq1e4}, a recently developed numerical-relativity-informed surrogate model based on black-hole perturbation theory framework. We validate our results against the remnant properties estimated from numerical relativity (NR) surrogate models in the comparable mass ratio regime and against recently available high-mass-ratio NR simulations at $q=[15,32,64]$.  We find that our remnant property estimates computed from fluxes at future null infinity closely match the estimates obtained from the NR surrogate model of apparent horizon data. Using Gaussian process regression fitting methods, we train a surrogate model, \texttt{BHPTNR\_Remnant}, for the properties of the remnant black hole arising from BBH mergers with mass ratios from $q=2.5$ to $q=1000$. Finally, we discuss potential improvements in the \texttt{BHPTNRSur1dq1e4} waveform model when including remnant information. We make both the \texttt{gw\_remnant} and \texttt{BHPTNR\_Remnant} packages publicly available.

\end{abstract}

\maketitle
%==========================================================================
%==========================================================================
%==========================================================================
\section{Introduction}
During the coalescence, binary black holes  (BBHs) dissipate energy, linear momentum, and angular momentum through gravitational radiation. As the system emits energy and angular momentum, the orbit of the binary shrinks, resulting in an inspiral and an eventual merger. At the time of the merger, black holes typically move at about half of the speed of light, making a BBH coalescence one of the most luminous events in the universe. Therefore, to accurately estimate the properties of the final black hole, it is essential to study the complete merger dynamics of the binary, starting from the early inspiral to ringdown.

The radiated energy corresponds to a mass deficit in the final black hole (often termed the remnant black hole), while loss of angular momentum impacts the spin of the final black hole. While gravitational radiation will always carry away energy and angular momenta, the binary must have some degree of asymmetry to dissipate linear momentum away. Most of the linear momentum is, however, dissipated away in the last few cycles of the binary evolution leading to a sudden recoil (or kick) on the binary's center of mass near the merger~\cite{PhysRev.128.2471,1973ApJ,Campanelli:2007cga,Gonzalez:2007hi,Lousto:2011kp}. 

Remnant properties play an important role throughout gravitational-wave science. 
For example, the ringdown spectra of the gravitational radiation are governed by the remnant black hole, which is characterized by the remnant mass, remnant spin, and the kick velocity~\cite{Israel:1967za,Carter:1971zc,varma2022evidence,ma2021universal}. Access to the remnant properties of the binary is therefore crucial to develop accurate models for the full inspiral-merger-ringdown signal. Accurate estimation of the remnant masses and spins are also necessary to perform null tests of general relativity using gravitational waves~\cite{Ghosh:2017gfp,LIGOScientific:2016lio} and probe certain binary formation channels where the remnant black hole may take part in second-generation mergers~\cite{sesana:2007zk,Gualandris:2007nm,Merritt:2004xa}.

Properties of the remnant black hole are typically computed either directly from local measurements of the remnant black hole's apparent horizon~\cite{boyle2019sxs} (possible in a numerical relativity simulation) or through radiated harmonic modes and invoking flux balancing arguments. These modes can be computed from inspiral-merger-ringdown (IMR) waveform models such as the effective-one-body models~\cite{bohe2017improved,cotesta2018enriching,cotesta2020frequency,pan2014inspiral,babak2017validating,Nagar:2022icd}, phenomenological models~\cite{husa2016frequency,khan2016frequency,london2018first,khan2019phenomenological}, or numerical-relativity surrogates~\cite{Blackman:2015pia,Blackman:2017pcm,Blackman:2017dfb,Varma:2018mmi,Varma:2019csw,Islam:2021mha,Yoo:2022erv}. While it is important to follow the fully relativistic nonlinear dynamics of the binary to accurately calculate the remnant properties, this process may be too slow depending on the time taken to generate the necessary data. To overcome this issue, many attempts have been made to develop accurate phenomenological fits to the remnant data by calibrating an analytical ansatz to a handful of NR data~\cite{Herrmann:2007ex,Campanelli:2007cga,Gonzalez:2007hi,Gonzalez:2006md,Campanelli:2007ew,Rezzolla:2007rz,Rezzolla:2007rd,Kesden:2008ga,Tichy:2008du,Lousto:2007db,Barausse:2009uz,Barausse:2012qz,Lousto:2012su,Lousto:2012gt,Healy:2014yta,Zlochower:2015wga,Hofmann:2016yih,Gerosa:2016sys,Healy:2016lce,Healy:2018swt,Sundararajan:2010sr,Gerosa:2018qay}. This method may be prone to systematic biases due to the choice of the ansatz. An alternative approach is to build data-driven surrogate models for the remnant's properties based on NR data~\cite{Varma:2018aht,Taylor:2020bmj,Jimenez-Forteza:2016oae}. While the resulting models have been shown to be both accurate and fast, these models are only valid in the comparable mass ratio regime where NR data is plentiful. Extending these remnant models beyond the comparable mass ratio regime has always been challenging due to the lack of NR simulations,
although remarkable progress has been made on this front~\cite{Lousto:2020tnb,Lousto:2022hoq,Giesler:2022inPrep}.

Recently, a numerical-relativity informed point-particle black-hole-perturbation theory (ppBHPT) based surrogate model, \bhpt{}~\cite{Islam:2022laz}, was developed for comparable-to-large mass ratio binaries~\cite{Rifat:2019ltp,Islam:2022laz}. The model was validated against a handful of NR waveforms for mass ratios ranging from $q=15$ to $q=32$. We will use this model to investigate the remnant properties of high-mass-ratio BBH mergers and to develop accurate remnant fits in this regime. The rest of the paper is organized as follows. In Sec.~\ref{sec:methods}, we present an overview of the methods used to compute the properties of the remnant black hole using the \bhpt{} model and provide a prescription to fit the remnant data. We then present our results in Sec.~\ref{sec:results} and discuss a possible implication for waveform modelling in the context of the \bhpt{} model in Sec.~\ref{sec:waveform_implication}. Finally, we discuss our results in Sec.~\ref{sec:discussion}.

%==========================================================================
%==========================================================================
%==========================================================================
\section{Methods}
\label{sec:methods}
In this section, we first provide an overview of our NR-informed ppBHPT waveform model (Sec.~\ref{sec:wf_model}). We then present an executive summary of the methods we use to compute the remnant properties of the binary given a gravitational waveform $h(t)$ (Sec.~\ref{sec:remnant_calculation}). Finally, we summarize the techniques used to build fits for the remnant properties as a function of the binary parameter space (Sec.~\ref{sec:remnant_fits}).

%==========================================================================
%==========================================================================
\subsection{Overview of the \texttt{BHPTNRSur1dq1e4} waveform model}
\label{sec:wf_model}
Gravitational radiation from the merger of a binary black hole is typically written as a superposition of $-2$ spin-weighted spherical harmonic modes with indices $(\ell,m$):
\begin{align}
h(t,\theta,\phi;\boldsymbol{\lambda}) &= \sum_{\ell=2}^\infty \sum_{m=-\ell}^{\ell} h^{\ell m}(t;\boldsymbol\lambda) \; _{-2}Y_{\ell m}(\theta,\phi)\,,
\label{hmodes}
\end{align}
where $\boldsymbol{\lambda}$ is the set of intrinsic parameters (such as the masses and spins of the binary) describing the system, $\theta$ is the polar angle, and $\phi$ is the azimuthal angle.

In this paper, we generate gravitational waveforms primarily using the \texttt{BHPTNRSur1dq1e4} model~\cite{Islam:2022laz}. The model can be accessed through \texttt{gwsurrogate}~\cite{
gwsurrogate,field2014fast} or through \texttt{BHPTNRSurrogate}~\cite{BHPTSurrogate} package from the \texttt{Black Hole Perturbation Theory Toolkit}~\cite{BHPToolkit}.
This is a surrogate model trained on waveform data generated by the ppBHPT framework for mass ratios varying from $q=2.5$ to $q=10^{4}$.  The full IMR ppBHPT waveform training data is computed using a time-domain Teukolsky equation solver, the details of which have appeared in the literature extensively ~\cite{Sundararajan:2007jg,Sundararajan:2008zm,Sundararajan:2010sr,Zenginoglu:2011zz,Islam:2022laz,Rifat:2019ltp}. The model includes a total of 50 spherical harmonic modes up to $\ell=10$, and calibrates these modes to numerical relativity data up to $\ell=5$ in the comparable mass regime ($2.5 \le q \le 10$).
In the comparable mass-ratio regime, including mass ratios as low as $2.5$, the gravitational waveforms generated through ppBHPT were found to agree surprisingly well with those from NR after this simple calibration step. For example, when compared to recent SXS and RIT NR simulations at mass ratios ranging from $q=15$ to $q=32$ the \texttt{BHPTNRSur1dq1e4}'s dominant quadrupolar mode agrees to better than $\approx 10^{-3}$. 

%==========================================================================
%==========================================================================
%==========================================================================
\subsection{Framework to compute the remnant properties}
\label{sec:remnant_calculation}
We compute remnant quantities (such as the remnant mass, remnant spin, remnant kick, and peak luminosity) from the gravitational waveform's harmonic modes~\cite{RevModPhys.52.299,Lousto:2007mh,Ruiz:2007yx,Gerosa:2018qay} mostly following the equations and conventions outlined in Ref.~\cite{Gerosa:2018qay}, which we provide for completeness.
We implement the framework in the \texttt{gw\_remnant}\footnote{\href{https://github.com/tousifislam/gw\_remnant}{https://github.com/tousifislam/gw\_remnant}} package and make it publicly available.

%==========================================================================
%==========================================================================
\subsubsection{Remnant mass}

The energy flux due to gravitational radiation is given by,
\begin{align}
\dot{E} = \lim_{r \rightarrow \infty} \frac{r^2}{16\,\pi}
\sum_{\ell, m} \,\left| \dot h^{\ell m} \right|^2 \; ,
\label{energyflux}
\end{align}
and where we use an over-dot to denote $\partial / \partial_t$. Integrating this expression, $E(t) = \int_{-\infty}^{t} \dot{E} (t') dt'$, gives the total radiated energy at time $t$. In many cases, we only have access to $h^{\ell m}$ over a finite duration and wish to know $E(t)$ for a hypothetical, quasi-circular BBH system that started from an infinitely large initial orbital separation. Let us define time such that $t=0$ occurs at the peak of the
total waveform amplitude (taken to be $\sum_{\ell m} \left| h^{\ell m} \right|$) and let $t_{\rm initial}$ be the start of the waveform modes we have access to. Then the total energy radiated by the hypothetical system at time $t$ can be written as
\begin{align}
E(t) = E_0 + \int_{t_{\rm initial}}^{t} \dot{E} (t') dt' \; ,
\label{energy}
\end{align}
where the constant $E_0$ accounts for the energy dissipated in GWs at times $t \leq t_{\rm initial}$. Let us assume that, at least in early inspiral, the system's binding energy decreases at a rate determined by energy flux due to gravitational radiation, a standard assumption in post-Newtonian (PN) models. We can then estimate $E_0$ using a PN expression for the binding energy  (see Eq.~2.35 of Ref.~\cite{LeTiec:2011ab})
\begin{equation}
\frac{E_{\rm PN}}{M} =  - \frac{1}{2} \nu x \left( 1 + E_{1}^{\rm PN} x + E_{2}^{\rm PN} x^2 + E_{3}^{\rm PN} x^3 \right),
\end{equation}
with
\begin{align}
E_{1}^{\rm PN} &= \left( -\frac{3}{4} - \frac{\nu}{12}\right) \; ,\\
E_{2}^{\rm PN} &=  \left( -\frac{27}{8} + \frac{19\nu}{8} - \frac{\nu^2}{24}\right) \; ,\\
E_{3}^{\rm PN} &=  \left( -\frac{675}{64} + \left[ \frac{34445}{576} - \frac{205\pi^2}{96}\right] \nu - \frac{155\nu^2}{96} - \frac{35\nu^3}{5185} \right) \; ,
\label{eq:E0}
\end{align}
where $\nu=\frac{q}{(1+q)^2}$ is the symmetric mass ratio,
$M=m_1 + m_2$ is the total mass,
$x=\omega^{1/3}$, $\omega=\frac{d\phi}{dt}$, and $\phi(t)$ is the orbital phase of the binary. 
The integration constant,
\begin{equation}
E_0 = - E_{\rm PN} (x(t_{\rm initial}))\,,
\label{eq:Mb}
\end{equation}
is then simply given by direct evaluation.

We now consider a model for the remnant mass.
Let $M_B(t)$ be the time-dependent Bondi mass of the binary system and $M_{\rm ADM}$ be the Arnowitt-Deser-Misner (ADM)~\cite{Arnowitt:1962hi} mass. For isolated systems, these two masses are related by~\cite{PhysRevLett.43.181},
\begin{equation}
M_{\rm ADM} = M_B(t) +  \int_{-\infty}^{t} \dot{E} (t') dt' \,.
\end{equation}
That is, the ADM mass equals the Bondi mass plus the energy carried away by gravitational radiation. The ADM mass can be directly computed in a numerical relativity simulation but is, of course, unavailable in waveform models. In App.~\ref{app:ADM}, we show that for the non-spinning systems considered here, the ADM mass as computed in a typical NR simulation can be approximated by~\footnote{This dependency can be understood by noting that there is no gravitational radiation content at the start of an NR simulation. Consequently, the ADM mass as computed in NR will depend only on the initial data (such as the initial coordinate separation of the two black holes) prescribed on the finite computational domain. For example, see Table 1 of Ref.~\cite{szilagyi2015approaching}.},
\begin{equation}
    M_{\rm ADM} \approx M - E_0 \,,
\end{equation}
where $E_0$ is the PN binding energy at the start of the waveform.
Assuming no gravitational wave emission at times $t \leq t_{\rm initial}$ (which is the case for NR simulations), our modelled Bondi mass is 
\begin{equation}
M_B(t) = M - E_0  -  \int_{t_{\rm initial}}^{t} \dot{E} (t') dt' \,,
\label{eq:Mb}
\end{equation}
from which the remnant mass,
\begin{align}
M_{\rm rem} &= M_B(\infty) \approx M_B(t_{\rm end}) \nonumber \\
& = M - E_0  - \int_{t_{\rm initial}}^{t_{\rm end}} \dot{E} (t') dt' \,,
\label{eq:Mrem}
\end{align}
is readily computed. Here $t_{\rm end}$ is time at the end of the waveform, set to be $t_{\rm end} \approx 115M$ in the \texttt{BHPTNRSur1dq1e4} model. Our remnant model and the assumptions underlying it are supported by Fig.~\ref{fig:remnant_q1to10} (top row), where we compare to the remnant mass computed directly from the final black hole's apparent horizon (solid red line).

%==========================================================================
%==========================================================================
\subsubsection{Remnant kick velocity}
The time derivative of the radiated linear momenta is expressed as:
\begin{align}
\frac{d P_x}{dt} = &\lim_{r \to \infty} \frac{r^2}{8\, \pi} \Re \Bigg[ \sum_{\ell,m} \,
\dot h^{\ell m}  \Big( a_{\ell m}\, \dot{\bar{h}}^{\ell,m+1}
 \notag\\
&+ b_{\ell,-m} \,\dot{\bar{h}}^{\ell-1,m+1}  -  b_{\ell+1,m+1}\, \dot{\bar{h}}^{\ell+1,m+1} \Big)\Bigg] \; ,
\label{eq:dt_px} \\
\frac{d P_y}{dt} = &\lim_{r \to \infty} \frac{r^2}{8\, \pi}\Im \Bigg[  \sum_{\ell,m}\, \dot h^{\ell m}  \Big( a_{\ell m}\, \dot{\bar{h}}^{\ell,m+1}
 \notag\\
&+ b_{\ell,-m} \,\dot{\bar{h}}^{\ell-1,m+1}  -  b_{\ell+1,m+1}\, \dot{\bar{h}}^{\ell+1,m+1} \Big)\Bigg] \; ,
\label{eq:dt_py} \\
\frac{d P_z}{dt} = &\lim_{r \to \infty} \frac{r^2}{16 \pi} \sum_{\ell,m}\,
\dot{{h}}^{\ell m}   \Big( c_{\ell m}\, \dot{\bar{h}}^{\ell m}
\notag\\
&+ d_{\ell m}\, \dot{\bar{h}}^{\ell-1,m} +  d_{\ell+1,m}\, \dot{\bar{h}}^{\ell+1,m} \Big) \; ,
\label{eq:dt_pz}
\end{align}
where the upper bar denotes complex conjugation and the coefficients $a_{\ell,m}$, $b_{\ell,m}$, $c_{\ell,m}$, and $d_{\ell,m}$ are given in Ref.~\cite{Gerosa:2018qay}.
Integrating $d\mathbf{P}/dt$ then gives us the radiated linear momentum of the binary as a function of time. While integrating, we set the integration constant to be zero as the linear momentum emission at early inspiral is expected to be negligible when averaged over an orbital cycle. The time-dependent kick
imparted to the system's center of mass is then,
\begin{equation}
\mathbf{v}(t) = - \frac{{P_x}(t) \mathbf{\hat x} + {P_y}(t) \mathbf{\hat y} + {P_z}(t) \mathbf{\hat z}}{M_B(t)}\,,
\label{voftprofile}
\end{equation}
and taking the magnitude
\begin{align}
v_{\rm rem}^{\rm kick} = |\mathbf{v}(t=t_{\rm end})|\,,
\label{vkicklimit}
\end{align}
gives the kick velocity of the remnant.

%==========================================================================
%==========================================================================
\subsubsection{Remnant spin}
The rate of loss of angular momentum during binary evolution has the following form:
\begin{align}
\frac{d J_x}{dt} =  &\lim_{r\rightarrow\infty} \frac{r^2}{32 \pi} \:
\Im \Bigg[ \sum_{\ell m} \,h^{\ell m}
\Big( f_{\ell m}\, \dot{\bar{h}}^{\ell,m+1}
\notag \\&+ f_{\ell,-m}\, \dot{\bar{h}}^{\ell,m-1} \Big) \Bigg]\;  ,
\label{eq:dt_jx} \\
\frac{d J_y}{dt} = &- \lim_{r\rightarrow\infty} \frac{r^2}{32 \pi} \:
\Re \Bigg[ \sum_{\ell,m} \, h^{\ell m}  \Big( f_{\ell m}\, \dot{\bar{h}}^{\ell,m+1}
\notag \\ &- f_{\ell,-m}\, \dot{\bar{h}}^{\ell,m-1} \Big) \Bigg]\; ,
\label{eq:dt_jy} \\
\frac{d J_z}{dt} =  &\lim_{r\rightarrow\infty} \frac{r^2}{16 \pi} \:
\Im \Bigg[ \sum_{\ell,m} \,m\, h^{\ell m}
 \,\dot{\bar{h}}^{\ell m} \Bigg] \; ,
\label{eq:dt_jz}
\end{align}
where
\begin{eqnarray}
f_{\ell m} = \sqrt{\ell(\ell+1) - m(m+1)} \; .
\label{flm}
\end{eqnarray}
To compute the total loss of the angular momentum since the start of the waveform, we integrate $d\mathbf{J}/dt$. 

For the non-spinning BBH systems considered here, the orbital motion  is confined to the x-y plane. Let $J^0_z$ be the orbital angular momentum at the start of the waveform, then invoking 
a flux balancing argument, we can estimate the remnant spin of the final black hole
by~\cite{radia2021anomalies}
\begin{eqnarray}
a_{\rm rem} = \frac{J^0_z-J_z^{\rm rad}}{M_{\rm rem}^2}\;,
\label{remnantspin}
\end{eqnarray}
where $J_z^{\rm rad}$ is the total loss of angular momentum since the start of the waveform and, due to the symmetry of the systems we consider, the final spin is always in the $z$ direction.
We estimate $J^0_z$ using a post-Newtonian expression for the angular momentum (see Eq.~2.36 of Ref.~\cite{LeTiec:2011ab})
\begin{equation}
\frac{J_{\rm PN}}{M^2} =  \frac{\nu}{x^{1/2}}\left( 1 + J_{1}^{\rm PN} x + J_{2}^{\rm PN} x^2 + J_{3}^{\rm PN} x^3 \right),
\end{equation}
where
\begin{align}
J_{1}^{\rm PN} &= \left( \frac{3}{2} + \frac{\nu}{6}\right) \; ,\\
J_{2}^{\rm PN} &=  \left( \frac{27}{8} - \frac{19\nu}{8} + \frac{\nu^2}{24}\right) \; ,\\
J_{3}^{\rm PN} &=  \left( \frac{135}{16} + \left[ -\frac{6889}{144} + \frac{41\pi^2}{24}\right] \nu + \frac{31\nu^2}{24} + \frac{7\nu^3}{1296} \right) \;.
\label{eq:J0}
\end{align}\\
The integration constant $J^0_z$ is then taken to be $J_{\rm PN}(x(t_{\rm initial}))$.

%==========================================================================
%==========================================================================
\subsubsection{Peak luminosity}

We calculate the peak value luminosity,
\begin{align}
\mathcal{L}_{\rm peak} = \max_{t} \dot{E} \; ,%\mathcal{L}(t).\;
\label{peakluminosity}
\end{align}
by fitting a quadratic function to 21 adjacent
samples of $ \dot{E}$, 
consisting of the largest sample and ten neighbors on either side. The peak luminosity can then be found analytically from
the fit.

%==========================================================================
%==========================================================================
\subsection{Building fits for the remnant properties}
\label{sec:remnant_fits}
One of the primary objectives of this paper is to provide accurate fits to the remnant properties
for both comparable and large mass ratio BBH mergers. We first compute the remnant properties -- remnant mass $M_{\rm rem}$, remnant spin $a_{\rm rem}$, 
remnant kick velocity $v_{\rm rem}^{\rm kick}$, and the peak luminosity $\mathcal{L}_{\rm peak}$ -- using gravitational waveforms generated with the \bhpt{} model. 

\begin{figure*}
\includegraphics[scale=0.43]{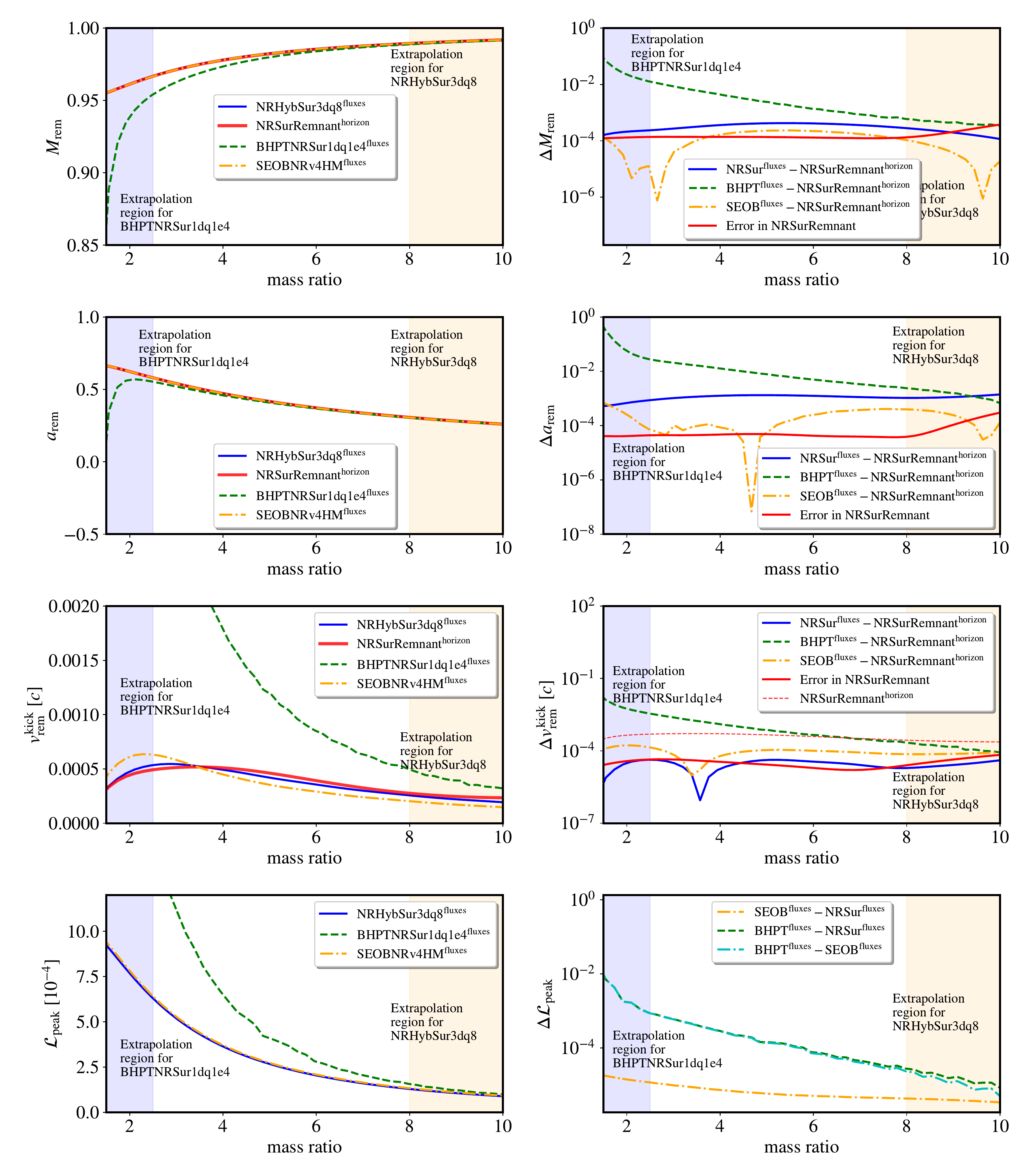}
\caption{Remnant quantities computed from three different waveform models \bhpt{} (dashed green), \seob{} (dash-dot yellow), and \nrsur{} (solid blue) as well as from \nrrem{} (solid red). The mass and spin values are computed either directly from apparent horizon measurements (\nrrem{}) or gravitational-wave fluxes (\bhpt{}, \seob{}, \nrsur{}). We show remnant mass $M_{\rm rem}$, remnant spin $a_{\rm rem}$, remnant kick velocity $v_{\rm rem}^{\rm kick}$ and the peak luminosity $\mathcal{L}_{\rm peak}$ in the left column and their respective errors in the right column. Shaded regions indicate the respective regions where \bhpt{} (blue) and \nrsur{} (yellow) models are extrapolated outside their training region, respectively. Details can be found in the text. To put kick velocity errors into perspective, we also show the \nrrem{} predictions as a dashed red line.
}
\label{fig:remnant_q1to10}
\end{figure*}

For the remnant mass, we choose to fit ${\rm log_{10}}(1-M_{\rm rem})$ instead of $M_{\rm rem}$ 
as it has a better-behaved functional form over a large range of mass ratios leading to more accurate fits. For the same reason, we build fits for ${\rm log_{10}}(a_{\rm rem})$, ${\rm log_{10}}(v_{\rm rem}^{\rm kick})$, and $\log_{10}{(\mathcal{L}_{\mathrm{peak}})}$. All of our fits are parameterized by ${\rm log_{10}}(q)$. To construct the fits, we use the Gaussian process regression (GPR) methods as implemented in \texttt{scikit-learn} with radial basis functions kernels. 
When the models are evaluated, we can easily get the predicted remnant by undoing these transformations.

\begin{figure*}
\includegraphics[scale=0.5]{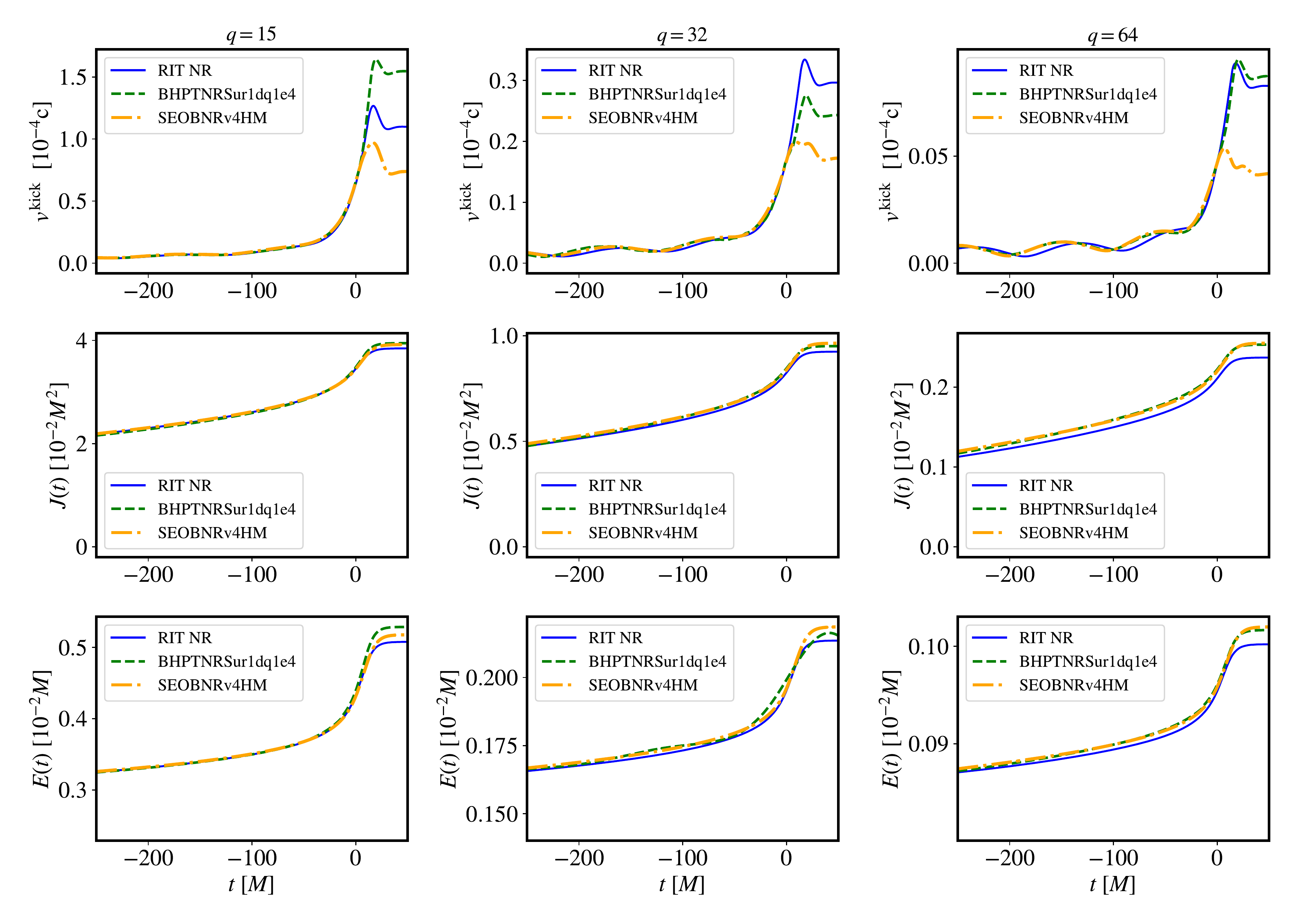}
\caption{We show the kick velocity profile $v_{\rm kick}$, the time-dependent radiated angular momentum profile $J(t)$ and the radiated energy profile $E(t)$ for three different mass ratio values: $q=15$ (left column), $q=32$ (middle column) and $q=64$ (right column). We show the profiles obtained from the RIT NR data (solid blue lines), the \bhpt{} model (green dashed lines), and the \seob{} model (orange dash-dotted lines).}
\label{fig:remnant_q15_32_64}
\end{figure*}

%==========================================================================
%==========================================================================
%==========================================================================
\section{Results}
\label{sec:results}
In this section, we first present a comparison between remnant properties computed using the \bhpt{} model and other state-of-art waveform models in the comparable mass ratio regime (Sec.~\ref{sec:result_q1to10}) and compare with RIT NR data in the intermediate mass ratio regime (Sec.~\ref{sec:result_RITNR}). Finally, we build GPR fits for the remnant quantities, obtained using \bhpt{}, for all mass ratios (Sec.~\ref{sec:result_fits}).

%==========================================================================
%==========================================================================
\subsection{Comparison against NR surrogates in the comparable mass ratio regime}
\label{sec:result_q1to10}
We first compute remnant quantities in the comparable mass ratio regime ($1 \le q \le 10$) using three different waveform models: \bhpt, \nrsur{}~\cite{Varma:2018mmi}  and \texttt{SEOBNRv4HM}~\cite{Cotesta:2020qhw}. \nrsur{} is a surrogate model for hybridized NR waveforms from aligned-spin BBH mergers trained on 104 NR waveforms for mass ratio $1\le q \le 8$ and spin $|\chi_1,\chi_2| \le 0.8$. The model can, however, be extrapolated
up to mass ratio $q \approx 10$. The model includes all spin-weighted spherical harmonic modes up to $\ell=4$ and the $(5,\pm5)$ but not the $(4,\pm1)$ or $(4,0)$ modes.  \texttt{SEOBNRv4HM} is a state-of-art effective-one-body model for the aligned-spin binaries and has the following four higher order modes apart for the dominant quadrupolar mode of radiation: $\{(\ell,m)=\{(2,\pm1),(3,\pm3),(4,\pm4),(5,\pm5)\}$. We use both \nrsur{} and \texttt{SEOBNRv4HM} in their non-spinning limit. For all models, we generate waveforms on the same time grid $t \in [-5000,100]M$ with time spacing $dt=0.1M$. To compute the remnant quantities, we use the following set of modes for both \bhpt{} and \nrsur{} models: $\{(\ell,m)=\{(2,\pm2),(2,\pm1),(3,\pm1),(3,\pm2),(3,\pm3),(4,\pm2), (4,\pm3)$ $,(4,\pm4)\}$. For \texttt{SEOBNRv4HM}, we use all available modes.

In Fig.~\ref{fig:remnant_q1to10}, we show the remnant mass, remnant spin, remnant kick velocity, and the peak luminosity $\mathcal{L}_{\rm peak}$
estimated from these three different waveform models (left column). For comparison, we also show remnant quantities computed from the \nrrem{} model~\cite{Varma:2018aht}. Notably, \nrrem{} predicts the remnant mass and spin values determined {\em directly} from the final black hole's apparent horizon~\cite{boyle2019sxs}, whereas we use fluxes computed from waveform modes where the integration constants are set from PN. 
A careful comparison of 
remnant properties from horizon data
and asymptotic data in numerical relativity simulations was recently reported on by Iozzo et al.~\cite{iozzo2021comparing}. 

We find that the remnant quantities obtained from waveforms generated with \nrsur{} model match closely to the values obtained from the \nrrem{} model, indicating the effectiveness of the assumptions underlying the framework to compute remnant quantities entirely from gravitational wave data (cf. Sec.~\ref{sec:remnant_calculation}). It is not surprising that the remnant estimates from the \bhpt{} waveforms (which are computed within perturbation theory) differ from the \nrrem{} estimates in the comparable mass regime. These differences quickly reduce as we increase the mass ratio, and for $q\ge 8$, remnant estimates from the \bhpt{} model become comparable to other waveform models used in this work. 
While we cannot make extensive comparisons to NR for $q \geq 10$, the \bhpt{} model (and hence remnant quantities computed from it) is expected to become more accurate in the high-mass-ratio regime where perturbation theory is more applicable.

%==========================================================================
%==========================================================================
\subsection{Validation against RIT NR data in the intermediate mass ratio regime}
\label{sec:result_RITNR}

To further check the remnant properties estimated from the \bhpt{} model, we validate our results against a handful of recently available NR simulations in the intermediate mass ratio range at $q=[15,32,64]$~\cite{Lousto:2022hoq}. Figure \ref{fig:remnant_q15_32_64} shows the kick velocity profile $v_{\rm kick}(t)$, the radiated angular momentum profile $J(t)$, and the radiated energy profile $E(t)$ computed with waveforms generated from the \bhpt{} model using the following set of modes: $\{(\ell,m)=\{(2,\pm2),(2,\pm1),(3,\pm1),(3,\pm2),(3,\pm3),(4,\pm2), (4,\pm3)$ $,(4,\pm4)\}$. We validate these results against the respective profiles estimated from the RIT NR data using the same set of modes. For further comparison, we also include remnant profiles estimated from the \seob{} waveforms using all modes available for that model. We note that the NR waveforms only provide the last $\sim2000M$ of the binary's evolution, so we use the same length of waveform data from \bhpt{} and \seob{} models. We find that the remnant profiles obtained from the \bhpt{} model are consistent with the NR data. While some small discrepancies are apparent in the figure, we note that the high-mass-ratio NR waveforms appear to have some residual eccentricity that could account for this.

\begin{figure*}
\includegraphics[scale=0.52]{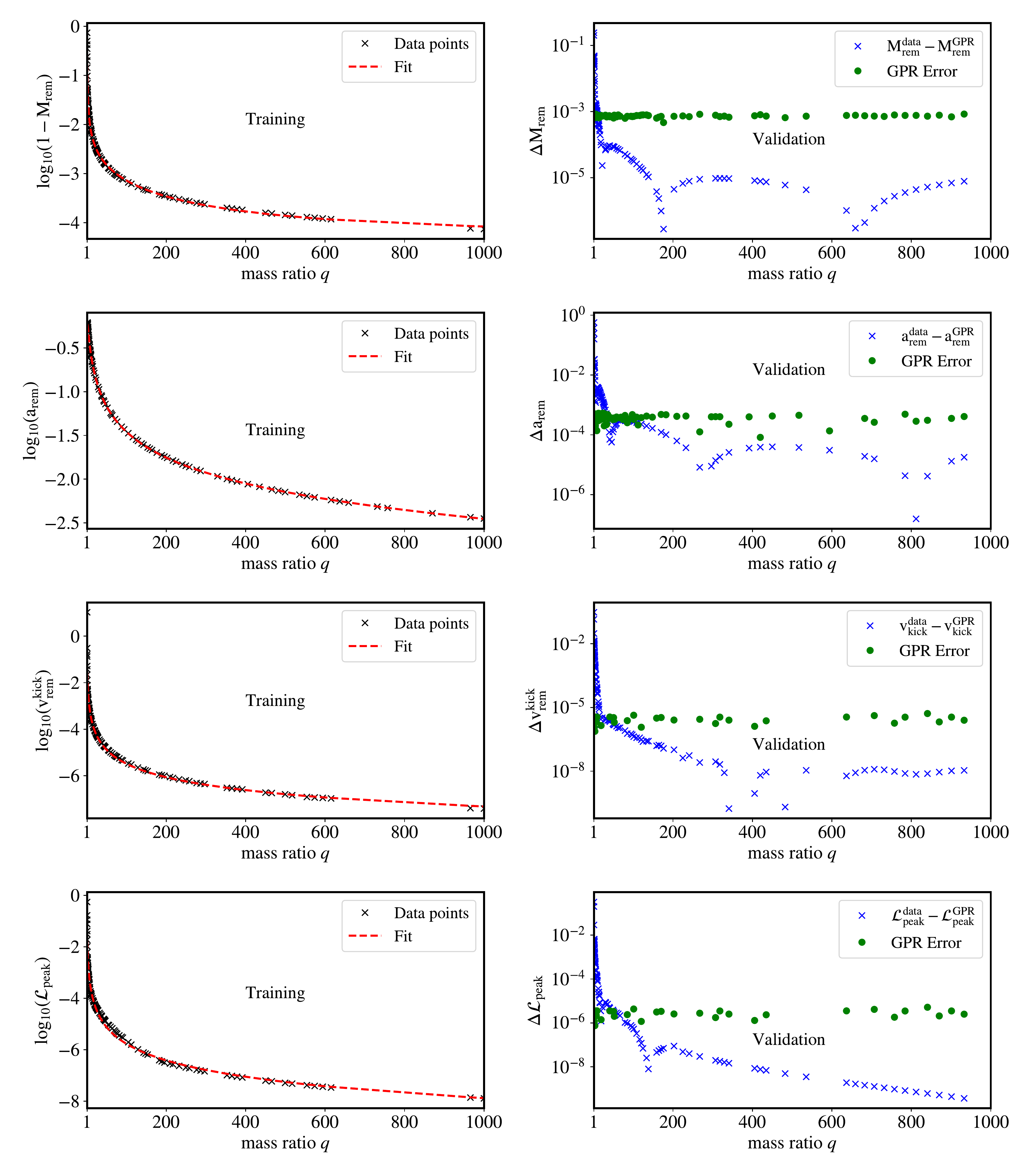}
\caption{\textit{Left column:} We show the training data (black crosses) and the GPR model's prediction (red dashed line) as a function of the mass ratio for $M_{\rm rem}$, $a_{\rm rem}$, $v_{\rm rem}^{\rm kick}$ and $\mathcal{L}_{\rm peak}$. \textit{Right panel: }we show the underlying GPR fit uncertainties (green circles) and the difference between the data and mean GPR predictions (blue crosses) in our validation set. Please note that the fit uncertainties (``GPR Error") are for the fitted data (e.g. $\log_{10}(1-M_{\rm rem})$) while the fit errors are for the remnant quantities (e.g. $M^{\rm data}_{\rm rem}-M^{\rm GPR}_{\rm rem}$)}.
\label{fig:fits}
\end{figure*}

%==========================================================================
%==========================================================================
\subsection{Fits for remnant properties for all mass ratios}
\label{sec:result_fits}
Results obtained in Secs.~\ref{sec:result_q1to10} and~\ref{sec:result_RITNR} demonstrate that the \bhpt{} waveform model can be used to predict the remnant properties in the intermediate mass ratio regime. Next, we build fits for the remnant mass, remnant spin, remnant kick velocity, and peak luminosity as a function of the mass ratio in the range $q \in [2,1000]$. We first choose $200$ different values of $q$ distributed uniformly in $\log_{10}(q)$. Following the framework described in Sec.~\ref{sec:remnant_calculation} and using the \bhpt{} model, we compute the remnant properties at these mass ratios. These remnant values will serve as both training and validation datasets.

\subsubsection{GPR fits}
We divide the remnant dataset into two separate groups using 100 data points for training and 100 data points for validation. Model fitting is performed using the procedure described in Sec.~\ref{sec:remnant_fits}. 

In Fig.~\ref{fig:fits} (\textit{left column}), we show both the training data as well as the GPR prediction (given by the GPR model's mean values) as a function of the parameter space. The right column compares the fit outputs against the validation data set and the estimated GPR fit uncertainties.
Based on the accuracy, we find that our fits are most useful at $q \geq 10$, which is sufficient for our purpose as other models (e.g. \nrrem{}) have been developed for mass ratios $q \leq 10$.
Our model complements these by working in the large-mass-ratio regime. We make our remnant fits publicly available through the \texttt{BHPTNR\_Remnant}\footnote{\href{https://github.com/tousifislam/BHPTNR_Remnant}{https://github.com/tousifislam/BHPTNR\_Remnant}} package.

\begin{figure}
\includegraphics[width=\columnwidth]{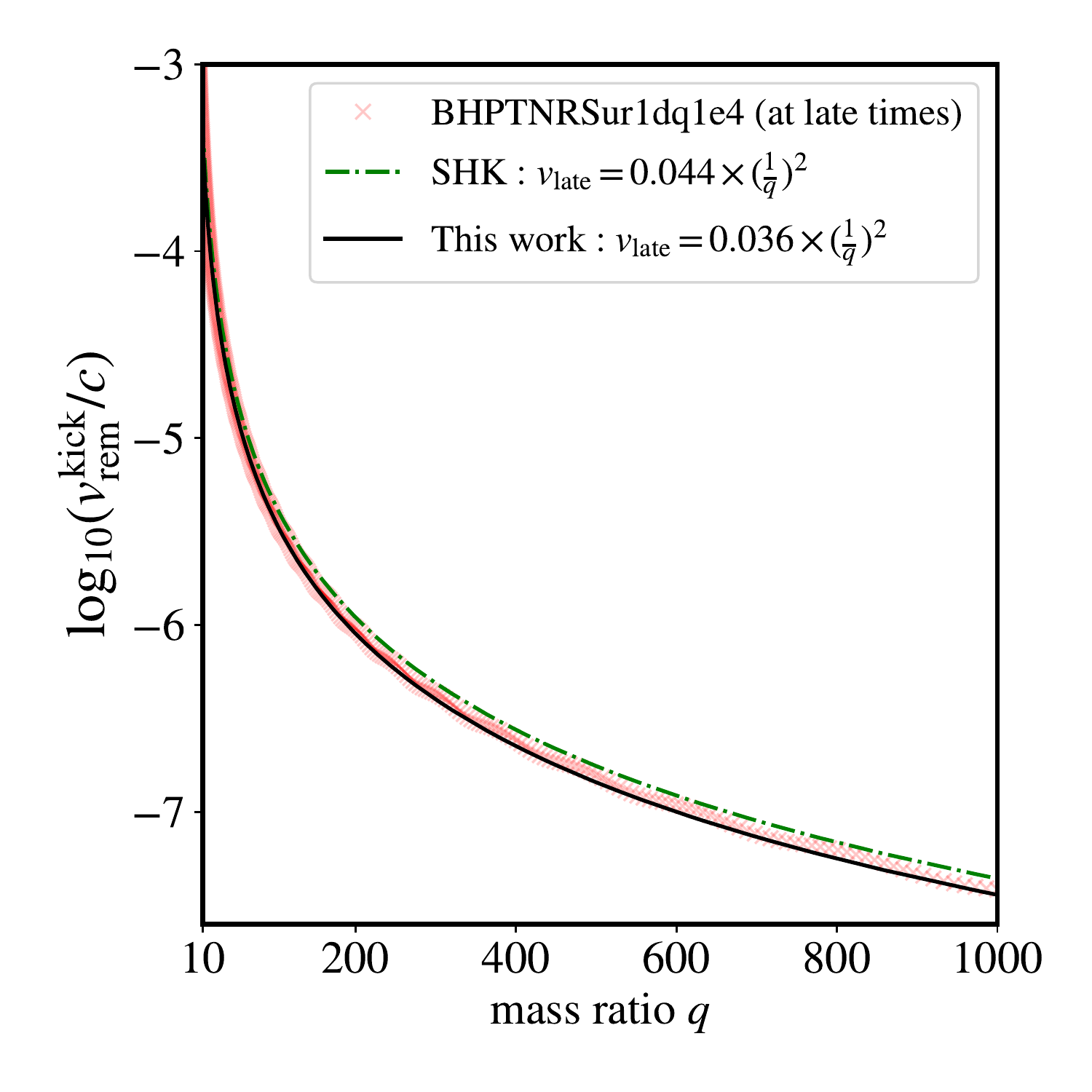}
\caption{We show the kick velocities $v_{\rm rem}^{\rm kick}$ (at late times) estimated from the \bhpt{} model (at late times) as a function of mass ratio $q$ (red crosses) along with the predictions from the fitted kick velocity formula from \texttt{SKH}~\cite{Sundararajan:2010sr} (green dash-dotted line). We also show the updated analytical fit in black. }
\label{fig:SKH_kick}
\end{figure}

\subsubsection{Analytical fit for the kick velocity}
As a final piece in our fitting exercise, we revisit the analytical kick velocity fits obtained by Sundararajan, Khanna, and Hughes (hereafter \texttt{SKH})~\cite{Sundararajan:2010sr}. They had modelled the kick velocity profile's peak,
\begin{equation}
v_{\rm kick}^{\rm peak}=0.051 \times (1/q)^2  \,,
\label{SKH:peak}
\end{equation}
and late-time (final) kick,
\begin{equation}
v_{\rm kick}^{\rm late}=0.044 \times (1/q)^2 \,. 
\label{SKH:late}
\end{equation}
In Fig.~\ref{fig:SKH_kick}, we show the remnant kick magnitudes (at late times) estimated from the \bhpt{} model as a function of the mass ratio $q$ along with the late time kick velocities obtained from \texttt{SKH} fits. We find that \texttt{SKH} fits result in a larger magnitude of kicks throughout the mass ratio range considered in this work. This is not surprising as the NR-informed \bhpt{} model yields a smaller amplitude than the raw ppBHPT waveforms used in obtaining \texttt{SKH} fits. While there could be additional explanations for discrepancies, the known differences in amplitude are likely to be one of the main reasons why the kick velocities obtained from the \bhpt{} model are systematically smaller than the values obtained from \texttt{SKH} fits. However, the functional form used in \texttt{SKH} fits catches the key behavior of the kick velocity. This suggests that one can possibly improve the \texttt{SKH} kick velocity model by updating the fit coefficients. Using \texttt{scipy.optimize.curve\_fit}, we obtain the following fitted formula for the kick velocity:
\begin{equation}
v_{\rm kick}^{\rm late}=0.034 \times (1/q)^2. 
\end{equation}
We find that the updated kick velocity fit matches the estimated kicks from the \bhpt{} model quite well for mass ratio $q\ge 5$. The fit accuracy further improves when we fit the data for $q\ge10$. In that case, we find the late-time and peak kick formula to be 
\begin{align}
v_{\rm kick}^{\rm late}& =0.036 \times (1/q)^2\, \\
v_{\rm kick}^{\rm peak}&=0.0401 \times (1/q)^2.
\end{align}

It has been previously argued in Ref.~\cite{Favata:2004wz} that one can improve the ability of ppBHPT to extrapolate out of the perturbative regime by replacing the $q^{-2}$ factor, which describes the momentum flux and the recoil velocity with $f(q)=\frac{1}{q^2} \sqrt{1-\frac{4}{q}}$. We verify that applying the $\frac{1}{q^2} \rightarrow f(q)$ rule on Eq.~(\ref{SKH:peak}) and Eq.~(\ref{SKH:late}) only changes the fit behavior in the small mass ratio regime (for $q\le 20$) while for larger mass ratios both fits give similar results.

\begin{figure}
\includegraphics[width=\columnwidth]{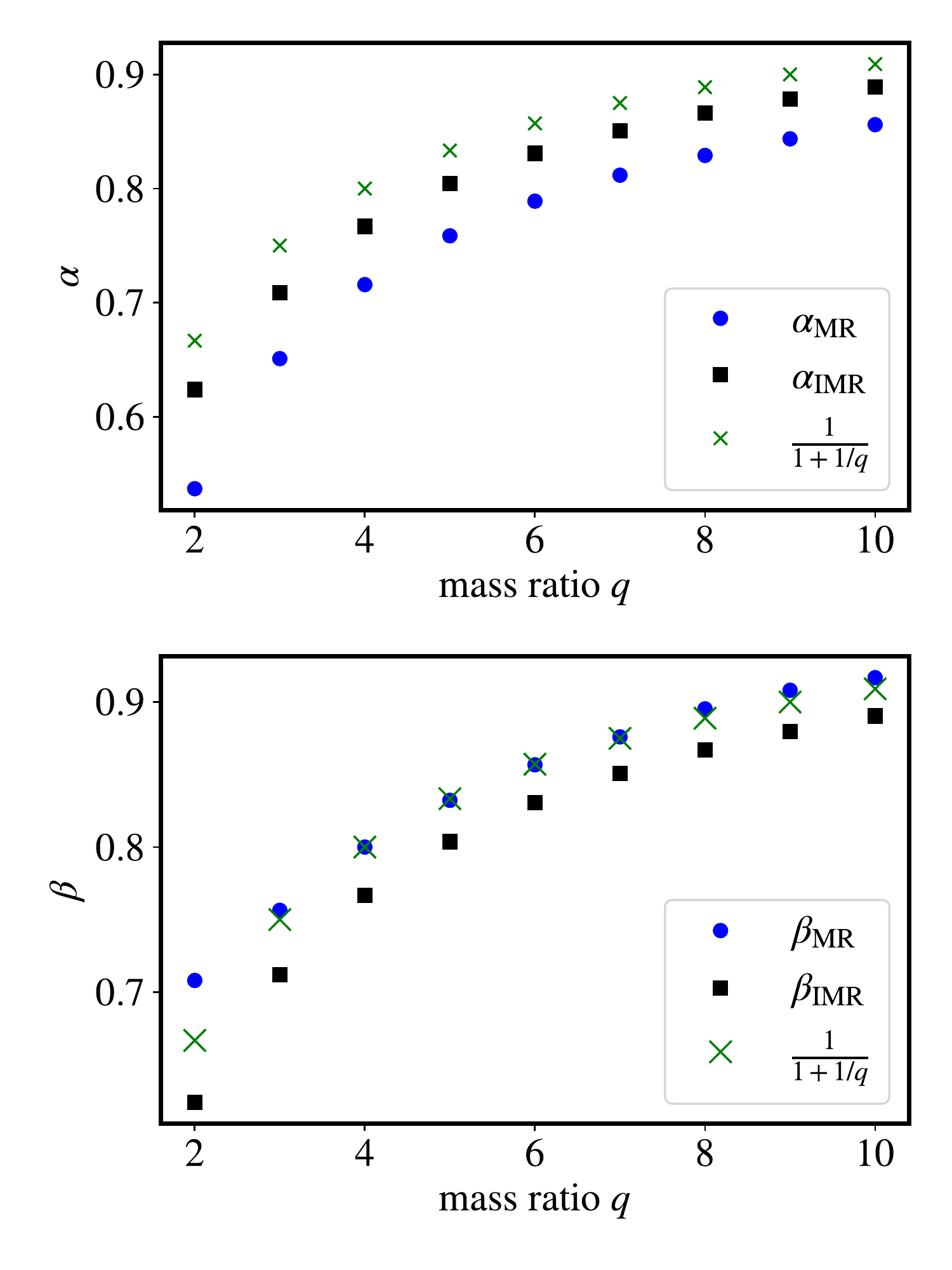}
\caption{We show the $\alpha$ (upper panel) and $\beta$ (upper panel) values used in the \bhpt{} model (denoted by $\alpha_{\rm IMR}$ and $\beta_{\rm IMR}$ respectively) as well as the ones computed using only the merger-ringdown part of the waveforms (denoted by $\alpha_{\rm MR}$ and $\beta_{\rm MR}$ respectively). For comparison, we also show the simple scaling factor $\frac{1}{1+1/q}$ needed to account for different definitions of the mass scale in NR and ppBHPT.
}
\label{fig:alpha_beta}
\end{figure}

%==========================================================================
%==========================================================================
%==========================================================================
\section{Application to waveform modelling}
\label{sec:waveform_implication}
\begin{figure}
\includegraphics[width=\columnwidth]{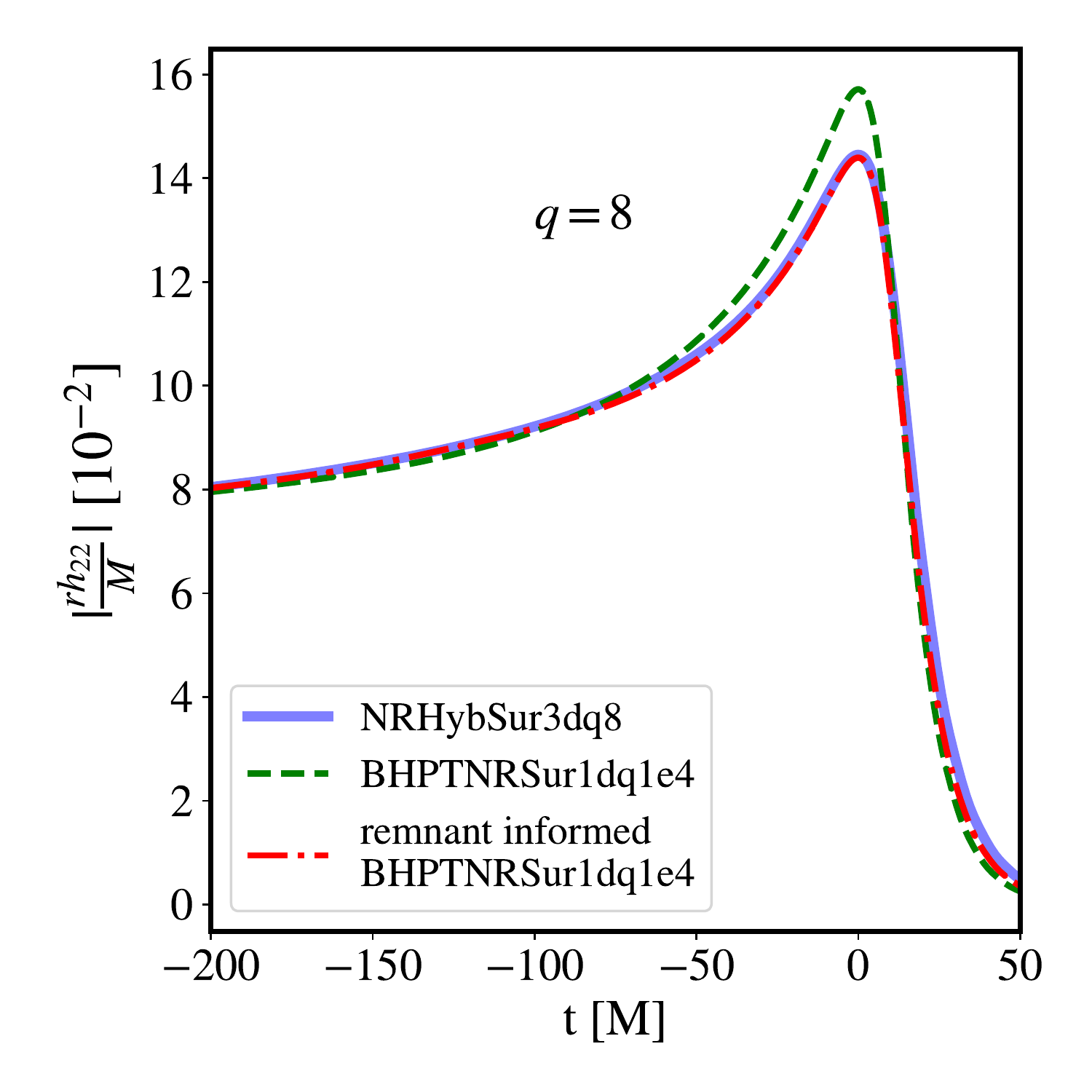}
\caption{We demonstrate how the remnant properties would help build better waveform models for the merger ringdown part of the waveform. We show the amplitude of the \bhpt{} waveform and the \nrsur{} waveform for $q=8$. For comparison, we then show the remnant informed \bhpt{} waveform too.
}
\label{fig:q8}
\end{figure}

We now explore whether the estimated remnant properties may help ongoing efforts to build waveform models based on the ppBHPT framework. For simplicity, we restrict ourselves to the $(2,2)$ mode. To better appreciate the problem, consider Fig.~\ref{fig:q8}, where we see noticeable disagreements between the peak amplitude of a $q=8$ waveform computed with \model{} (dashed green) and NR (solid blue). One explanation for this disagreement could be that within the ppBHPT framework, the remnant's value is equal to the mass of the primary, which is not physically correct. We would like to understand whether access to the remnant black hole's properties will help improve the \model{}'s merger and ringdown signal at comparable to intermediate-mass ratios.

We first recall that while the \model{} model is based on perturbation theory, the waveforms are calibrated to NR according to the formula:
\begin{align} \label{eq:EMRI_rescale}
h^{22}_{\tt BHPTNRSur1dq1e4, \alpha, \beta}(t ; q)= \alpha h^{22}_{\tt ppBHPT}\left( t \beta;q \right) \,,
\end{align}
where the dominant quadrupole, $h^{22}_{\tt ppBHPT}$, is computed using a high-order Teukolsky equation solver. Optimal values of $\alpha$ and $\beta$ are obtained by minimizing the difference
\begin{align} \label{eq:alpha_lm}
\min_{\alpha, \beta} \int \left| h^{22}_{\tt BHPTNRSur1dq1e4, \alpha, \beta}(t ; q) - h^{22}_{\rm NRHyb}(t ; q) \right|^2 dt \,,
\end{align}
between our model \model{} and a hybridized NR surrogate waveform \texttt{NRHybSur3dq8}~\cite{Varma:2018mmi} in 
its non-spinning limit 
over the time window $t\in [-5000,115]M$.
Because the calibration is performed using the full IMR waveform, our $\alpha$-$\beta$ calibration procedure is heavily influenced by the inspiral portion of the waveform. The merger-ringdown portion of the signal will, of course, be controlled by the remnant properties, which has not been taken into account. 
For example, we expect the binary to shed mass and angular momentum as it advances towards merger, which will change the waveform's post-merger signal by decreasing the amplitude.
One simple approach within our modelization setup is to allow the value of $\alpha$ and $\beta$ to vary between the inspiral and merger-ringdown part of the waveform. 
While this method will result in better matches to NR data in both the inspiral and the merger-ringdown part, it will also increase the number of free parameters by a factor of two.

\begin{figure}
\includegraphics[width=\columnwidth]{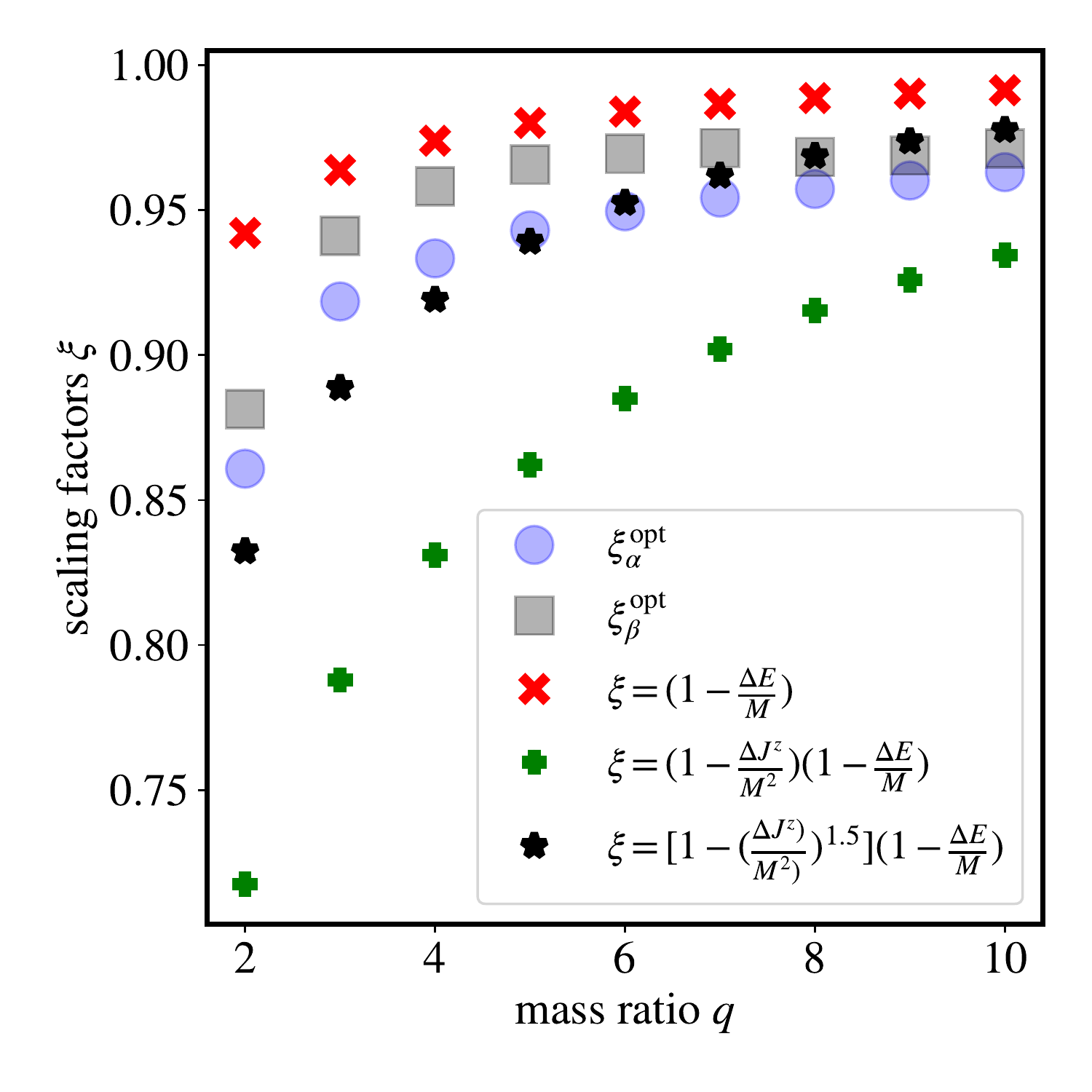}
\caption{We show the optimal scaling factors $\xi_{\alpha}^{\rm opt}$ (blue circles) and $\xi_{\beta}^{\rm opt}$ (gray squares) needed to take the $\alpha_{IMR}$ and $\beta_{\rm IMR}$ (scaling factors stemming from inspiral-merger-ringdown waveform data) to $\alpha_{\rm MR}^{\rm opt}$ and $\beta_{\rm MR}^{\rm opt}$ (scaling factors stemming from merger-ringdown waveform data only), respectively (see Fig.~\ref{fig:alpha_beta} and Eq.~\ref{eq:xi_def}). We also show three different proposed functional forms of $\xi$ for comparison.}
\label{fig:norm}
\end{figure}

Let us consider physically-motivated scaling factors $\{\xi_{\alpha}, \xi_{\beta} \}$ that will re-scale the already calibrated ppBHPT waveform in the post-merger signal according to
\begin{align} \label{eq:xi_def}
\alpha_{\rm MR} = \hspace{1mm}\xi_{\alpha} \times \alpha_{\rm IMR} \; ,  \qquad
\beta_{\rm MR} = \hspace{1mm}\frac{\beta_{\rm IMR}}{\xi_{\beta}} \; ,
\end{align}
where $\alpha_{\rm IMR}$ and $\beta_{\rm IMR}$ are the original calibration parameters used to build the \bhpt{} model (and are mostly influenced by the inspiral part of the waveform) and $\alpha_{\rm MR}$ and $\beta_{\rm MR}$ are new calibration parameters for the merger-ringdown waveform. This approach will keep the number of free parameters in our model unchanged if $\{\xi_{\alpha}, \xi_{\beta} \}$ can be determined without any extra free parameters. An obvious choice is to assume they are functions of the remnant mass and spin or, in a similar spirit, the radiated energy and angular momentum.

By solving the relevant optimization problem, we first compute optimal values of $\alpha_{\rm MR}^{\rm opt}$ and $\beta_{\rm MR}^{\rm opt}$
for a set of different mass ratios in the comparable mass regime, $2 \leq q \leq 10$, and plot their behavior in Fig.~\ref{fig:alpha_beta}. For comparison, we also plot $\alpha_{\rm IMR}$ and $\beta_{\rm IMR}$ as well as the scaling factor $\alpha = \beta = \left(1+1/q\right)^{-1}$ that, if used in Eq.~\eqref{eq:EMRI_rescale}, would account for the mass scale difference between ppBHPT (the mass scale is taken to be the primary $m_1$) and NR (the mass scale is taken to be the total mass $m_1 + m_2$ for non-spinning systems); see Refs.~\cite{Rifat:2019ltp,Islam:2022laz} for further discussion.

We now experiment with different functional forms for $\{\xi_{\alpha}, \xi_{\beta}\}$. We first compute the optimal factors, $\xi_{\alpha}^{\rm opt}$ and $\xi_{\beta}^{\rm opt}$, from $\alpha_{\rm MR}^{\rm opt}$, $\beta_{\rm MR}^{\rm opt}$, $\alpha_{\rm IMR}$, and $\beta_{\rm IMR}$. In Fig.~\ref{fig:norm}, we show $\xi_{\alpha}^{\rm opt}$ and $\xi_{\beta}^{\rm opt}$ as a function of the mass ratio. We find that the values for $\xi_{\alpha}^{\rm opt}$ and $\xi_{\beta}^{\rm opt}$ are quite close to each other. For simplicity of demonstration, we fix $\xi_{\alpha}=\xi_{\beta}=\xi$ and consider three possible choices for the scaling factor:
\begin{itemize}
    \item $\xi=(1-\frac{\Delta E}{M})$
    \item $\xi=(1-\frac{\Delta J^z}{M^2}) (1-\frac{\Delta E}{M})$
    \item $\xi=[1-(\frac{\Delta J^z}{M^2})^{1.5}] (1-\frac{\Delta E}{M})$
\end{itemize}
where $\Delta E$ and $\Delta J^z$ are the total radiated energy and total radiated angular momentum until $t_{\rm ref}=-100M$. The choice of $t_{\rm ref}=-100M$ is motivated by the intuition that the mass and the spin of the binary at the end of the plunge should characterize the merger-ringdown signal. We explored different choices of $t_{\rm ref}$ and found $t_{\rm ref}=-100M$ to consistently provide a better agreement between the post-merger signals.
In other words, $\alpha_{\rm MR}$ and $\beta_{\rm MR}$ should be related to the total loss of energy and angular momentum at plunge. Figure~\ref{fig:norm} shows the behavior of each choice of $\xi$. 

We now demonstrate a possible way of incorporating the remnant information into the
\bhpt{} model.
In Fig.~\ref{fig:q8}, we show the amplitude of the dominant $(2,2)$ mode of a $q=8$ waveform in the merger-ringdown part. We see a noticeable difference between the \bhpt{} model and \nrsur{} model. In particular, \bhpt{} predicts a larger amplitude. We employ the following procedure to incorporate the remnant information into the model. We first evaluate the \bhpt{} harmonic modes, from which the remnant properties can be computed.  We then re-scale the post-merger waveform using the scaling factor $\xi=(1-\frac{\Delta J^z}{M^2}) (1-\frac{\Delta E}{M})$. Finally, we hybridize the inspiral and post-merger parts using a smooth partition of unity.

We find that the remnant-informed \bhpt{} model matches the NR data much better than the usual \bhpt{} model. While the overall $L_2$-norm error, computed over the full IMR signal, only improves from $5.8 \times 10^{-4}$ to $4.6 \times 10^{-4}$, the improvement in the merger-ringdown part is much larger as clearly seen from Fig.~\ref{fig:q8}. We also find that the prescription works reasonably well for the higher modes too. 
While this simple example demonstrates the possible benefits of a remnant-informed \bhpt{}, achieving highly-accurate post-merger matches across all harmonic modes will be taken up in future work.

%==========================================================================
%==========================================================================
%==========================================================================
\section{Discussion \& Conclusion}
\label{sec:discussion}
The merger of two black holes results in a boosted Kerr black hole that is often referred to as the remnant.
In this paper, we implement a framework to compute the final mass, final spin, and kick of the remnant black hole using gravitational waveform modes (to compute fluxes) and post-Newtonian formula to set the relevant integration constants. We make this framework publicly available through the \texttt{gw\_remnant} Python package. 

Using the \texttt{gw\_remnant} package, we compute remnant quantities from the waveforms generated with the \bhpt{} model, comparing with  the \nrsur{} and \seob{} models as well as RIT NR waveform data at $q=[15,32,64]$. We find that remnant quantities found from the \bhpt{} model match these other models at mass ratios $q\ge 8$, which is to be expected as \bhpt{} is based on black hole perturbation theory and, therefore, is most applicable for intermediate- to large-mass-ratio systems. We also compare flux-based remnant quantities to the ones found from the \nrrem{} model that relies on quasi-local measurements on the remnant's apparent horizon to compute the final mass, spin, and kick. This provides a non-trivial comparison between remnant properties from waveform data and apparent horizon measurements performed in NR simulations. 

We also build surrogate models for the remnant properties of non-spinning BBH mergers as a function of mass ratios ranging from $q=2.5$ to $q=1000$, providing the first remnant models that can be used in the large-mass-ratio regime. As a byproduct of these studies, we also update previous analytical fits for the kick velocities. Such models for the remnant properties may help in developing efficient binary population models, probing high mass ratio BBH mergers, and investigating the nature of the remnant black hole using GW data. Our surrogate model is publicly available through the \texttt{BHPTNR\_Remnant} Python package.

Finally, we demonstrate how remnant quantities may help in building future NR-calibrated ppBHPT models, particularly in the merger-ringdown regime. Developing a robust framework to incorporate remnant information into the waveform modelling pipeline will be considered in future work.

%==========================================================================
%==========================================================================
%==========================================================================

\begin{acknowledgments}
The authors would like to thank Estuti Shukla for initiating the calculation of remnant properties from ppBHPT waveform data a few years ago (2020). The authors also acknowledge support from NSF Grants No.~PHY-2106755 (G.K), No. PHY-2110496 (T.I.~and S.F), and DMS-1912716 (T.I., S.F, and G.K). Part of this work is additionally supported by the Heising-Simons Foundation, the Simons Foundation, and NSF Grants Nos. PHY-1748958.
Simulations were performed on CARNiE at the Center for Scientific Computing and Visualization Research (CSCVR) of UMassD, which is supported by the ONR/DURIP Grant No.\ N00014181255, the UMass-URI UNITY supercomputer supported by the Massachusetts Green High Performance Computing Center (MGHPCC) and ORNL SUMMIT under allocation AST166. 
\end{acknowledgments}  

\appendix
\section{Approximating the ADM mass} 
\label{app:ADM}

To compute the remnant mass of a nonspinning binary according to Eq.~\eqref{eq:Mrem}, it is important to calculate the ADM mass of the spacetime. The ADM mass is readily computed in a numerical relativity simulation but unavailable in waveform models. In this appendix, we provide numerical evidence that the ADM mass, as computed in SpEC simulations~\cite{boyle2019sxs}, can be approximated as
\begin{equation}
    M_{\rm ADM} \sim M + E_{\rm binding}(t_{\rm initial}),
\end{equation}
where $E_{\rm binding}(=-E_0)$ is the post-Newtonian binding energy of the binary at the start of the waveform and $M=m_1 + m_2$  ($M$ is set to unity in a typical NR simulation). This dependency can be understood by noting that there is no gravitational radiation content at the start of an NR simulation. Consequently, the ADM mass as computed in NR will depend only on the initial data (such as the initial coordinate separation of the two black holes) prescribed on the finite computational domain.

\begin{figure}
\includegraphics[width=\columnwidth]{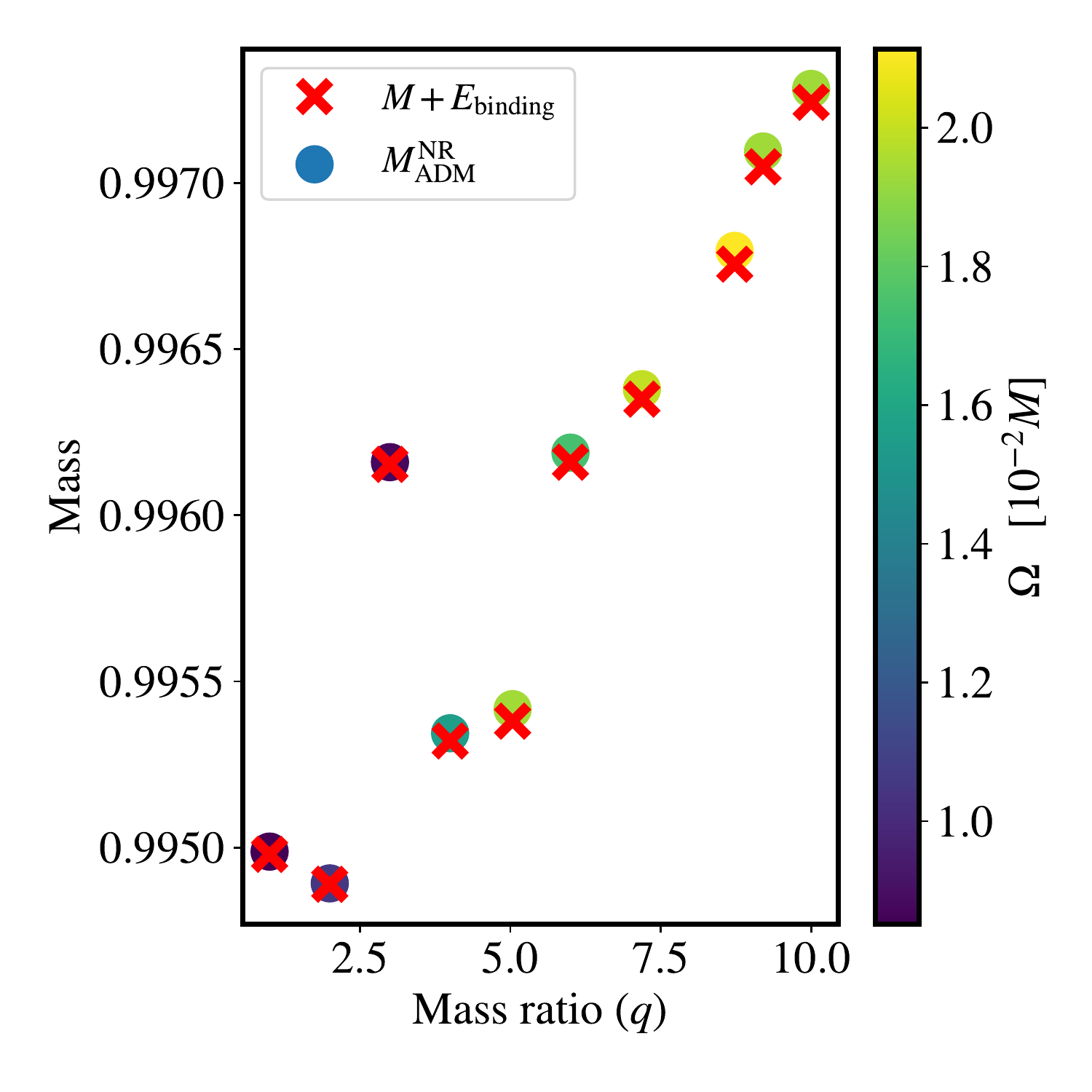}
\caption{We demonstrate that the ADM mass of the binary, $M_{\rm ADM}^{\rm NR}$, measured in an NR simulation (circles) can be well approximated by $M + E_{\rm binding}(t=t_{\rm initial})$ (red crosses). The color bar shows the value of the initial orbital frequency, $\Omega$. We see worsening agreement at larger values of $\Omega$, which is where PN approximations become less reliable. Here $t_{\rm initial}$ denotes that we evaluate the PN expression for the binding energy~\eqref{eq:E0} at the start of the simulation.
}
\label{fig:ADM}
\end{figure}

To demonstrate this, we select a set of 10 NR simulations for non-spinning BBHs from the SXS catalog~\cite{boyle2019sxs}~\footnote{\href{https://data.black-holes.org/waveforms/catalog.html}{https://data.black-holes.org/waveforms/catalog.html}}. For each NR simulation, we extract $M_{\rm ADM}$ from the simulation's metadata file. We then use the initial orbital frequency of the simulation (also taken from the metadata file) to compute $E_{\rm binding}$ using a PN expression~\eqref{eq:E0}. In Fig.~\ref{fig:ADM}, we show $M_{\rm ADM}$ as well as its approximated value $M + E_{\rm binding}$. We find that $M_{\rm ADM}$ and $M + E_{\rm binding}$ closely match with each other. Small discrepancies appear at larger values of initial orbital frequency, $\Omega$, which is where PN approximations become less reliable. Additional (probably less important) discrepancies may be due to junk radiation content lurking in the initial data and evaluation of the ADM integral on a finite outer boundary~\cite{lovelace2009reducing}.

Considering Fig.~\ref{fig:ADM}'s apparent dependence on $q$, we note that due to the computational challenges associated with high-mass-ratio NR simulations, available NR waveform durations vary with mass ratio. For example, the $q=1$ simulation covers $\sim25,000M$ in duration and has the lowest initial orbital frequency, whereas the $q=8.72$ simulation only covers $\sim5000M$ in duration and has the highest initial orbital frequency. 
When estimating the remnant mass according to Eq.~\eqref{eq:Mrem}, one would use a sufficiently long waveform such that the post-Newtonian binding energy estimated at $t_{\rm initial}$ is accurate.

%%%%%%%%%%%%%%%%%%%%%%%%%%%%%%%%%%%%%%%%%%%%%%%%%%%%%%%%%%%%%%%%%%%%%%%%%%%%%%%
\bibliography{BHPT_Remnant}
%%%%%%%%%%%%%%%%%%%%%%%%%%%%%%%%%%%%%%%%%%%%%%%%%%%%%%%%%%%%%%%%%%%%%%%%%%%%%%%

\end{document}